\documentclass[prd, twocolumn, amsfonts, amssymb, amsmath, showkeys, nofootinbib, superscriptaddress]{revtex4-2}

\usepackage[english]{babel}
\usepackage[utf8]{inputenc}
\usepackage{amsthm}
\usepackage{mathtools}
\usepackage{physics}
\usepackage{graphicx}
\usepackage{dcolumn}
\usepackage{bm}
\usepackage{xcolor}
\usepackage[pdftex, pdftitle={Article}, pdfauthor={Author}]{hyperref}
\usepackage{multirow}
\usepackage{hhline}
\usepackage{aas_macros}

\newcommand{\msun}{\mathrm{\, M_\odot}}
\newcommand{\chiNS}{\chi_{\mathrm{NS}}}
\newcommand{\chiBH}{\chi_{\mathrm{BH}}}
\newcolumntype{P}[1]{>{\centering\arraybackslash}p{#1}}

\begin{document}

\preprint{APS/123-QED}

\title{Inferring small neutron star spins with neutron star-black hole mergers}

\author{Ish Gupta}
\email[Correspondence email address: ]{ishgupta@psu.edu}
\affiliation{Institute for Gravitation and the Cosmos, Department of Physics, Pennsylvania State University, University Park, PA 16802, USA}



\begin{abstract}
The precise measurement of neutron star (NS) spins can provide important insight into the formation and evolution of compact binaries containing NS. While traditional methods of NS spin measurement rely on pulsar observations, gravitational wave detections offer a complementary avenue. However, determining component spins with gravitational waves is hindered by the small dimensionless spins of the NS and the degeneracy in the mass and spin parameters. This degeneracy can be addressed by the inclusion of higher-order modes in the waveform, which are important for systems with unequal masses. This study shows the suitability of neutron star-black hole mergers, which are naturally mass-asymmetric, for precise NS spin measurements. We explore the effects of the black hole masses and spins, higher-mode content, inclination angle, and detector sensitivity on the measurement of NS spin. We find that networks with next-generation observatories like the Cosmic Explorer and the Einstein Telescope can distinguish NS dimensionless spin of 0.04 (0.1) from zero at $1-\sigma$ confidence for events within $\sim 350$ $(\sim 1000)$ Mpc. Networks with A+ and A$^{\sharp}$ detectors achieve similar distinction within $\sim 30$ $(\sim 70)$ Mpc and $\sim 50$ $(\sim 110)$ Mpc, respectively.
\end{abstract}

\maketitle

\section{Introduction} \label{sec:intro}
The current ground-based gravitational wave (GW) detectors are sensitive to GW from binaries containing black holes (BH) and/or neutron stars (NS). These compact objects form as the end-product of the evolution of massive stars. The formation mechanism of such compact binaries can be broadly classified into two types--- first, where two closely-located massive stars evolve in isolation \cite{1973NInfo..27...70T,Postnov:2014tza}, and second, where the binaries are formed due to dynamical interactions in dense stellar environments \cite{Sigurdsson:1993zrm,Benacquista:2011kv}. The pathway followed for the binary formation and the processes it undergoes during its evolution leave their imprints on the system, particularly on the masses and spins of the components. Fortunately, by detecting the GWs from these systems, these parameters can be inferred, which provides important insights into the formation and evolution of the binary (e.g., see Ref. \cite{Mastrogiovanni:2022ykr}). As different formation channels lead to different parameters of the components \cite{LIGOScientific:2016vpg, Gerosa:2017kvu, Mandel:2018hfr}, and a single formation channel cannot explain the properties of the detected binaries \cite{Zevin:2020gbd}, precise estimation of the mass and the spin distributions is pivotal for understanding the contributions of the various formation channels to the binary population.

The first direct detection of GW was from a $\sim65\msun$, nearly equal mass, binary black hole (BBH) merger \cite{LIGOScientific:2016aoc}. This detection proved that BBH can merge in Hubble time and allowed the determination of the merger rate for such binaries, rejecting pessimistic formation scenarios that predicted very low merger rate values \cite{2014A&A...564A.134M,Belczynski:2006zi}. While the components' masses were more than the expectations from low-mass X-ray binary systems \cite{Farr:2010tu}, they were found to be consistent with both isolated and dynamical formation channels, provided that the corresponding metallicity is lower than solar metallicity \cite{LIGOScientific:2016vpg,Belczynski:2016obo}. Since then, the Advanced Laser Interferometer Gravitational-Wave Observatory (aLIGO) \cite{KAGRA:2013rdx,LIGOScientific:2014pky,aLIGO:2020wna,Tse:2019wcy} and the Advanced Virgo (AdV) \cite{VIRGO:2014yos,Virgo:2019juy} detector have detected $\sim100$ mergers by the end of the third observing run \cite{KAGRA:2021vkt}. Among these, the majority are BBH mergers, followed by two neutron star-black hole (NSBH\footnote{In literature, mergers with BH and NS are often referred to as NSBH if the NS forms first, and BHNS if the BH forms first. However, we use NSBH to refer to all mergers that contain an NS and a BH.}) mergers and two binary neutron star (BNS) mergers. Using the detected BBH mergers, significant constraints have been placed on the mass spectrum of BH \cite{KAGRA:2021duu}, revealing distinctive features that could contain information about the astrophysical processes at play \cite{Zevin:2017evb,Tiwari:2020otp,Wong:2020ise,Tiwari:2021yvr,Farah:2023vsc}. While the predictions for the masses of the components from different formation channels can overlap, the spin values and orientations can encode decisive information for distinguishing between the formation scenarios \cite{Vitale:2015tea,Farr:2017uvj,Farr:2017gtv}. The isolated binary formation channels are, predominantly, expected to create slowly spinning binaries with spins aligned with the orbital angular momentum \cite{Tutukov:1993bse,Belczynski:2016obo,Zaldarriaga:2017qkw,Fuller:2019sxi,Zevin:2022wrw} (however, see Refs. \cite{Kalogera:1999tq,OShaughnessy:2017eks,Olejak:2021iux}). On the other hand, binaries formed through dynamical exchanges are expected to have components with spins that are oriented isotropically \cite{Rodriguez:2016vmx} and possibly with large magnitudes \cite{Berti:2008af}. Unfortunately, as the individual spins of the compact objects cannot yet be precisely measured using GW observations, the spin distribution of the BHs from the current observations remains uncertain \cite{Vitale:2022dpa}, leading to a lack of clarity about the dominant formation channel for the detected population.

The properties of the cosmic population of BNS and NSBH mergers are even more uncertain, owing to the limited number of GW observations. Currently, our expectations are informed by the Galactic pulsar observations which point to a bimodal mass distribution with the dominant peak at $\sim1.4\msun$ and the minor peak at $\sim1.8\msun$ \cite{Antoniadis:2016hxz,Alsing:2017bbc,2020RNAAS...4...65F}. NS present in binaries are seen to coincide with the former peak \cite{Ozel:2012ax,Kiziltan:2013oja}. While GW170817 \cite{LIGOScientific:2017vwq} seems to follow this trend as well, the component masses inferred for GW190425 \cite{LIGOScientific:2020aai} contradict it. Unlike the pulsar population, the NS in BNS and NSBH mergers detected using GWs so far are consistent with a uniform distribution in mass \cite{Landry:2021hvl,KAGRA:2021duu}. 

Measuring the spin of the individual NS using GW is trickier. This is due to the following two reasons. Firstly, individual spins are difficult to measure. The GW signal for low-mass compact objects, like BNS and NSBH, is dominated by the inspiral--- where the orbit of the binary shrinks due to the emission of GW. This part of the signal is well approximated with the post-Newtonian (PN) formalism \cite{Blanchet:2013haa}, where the dominant spin effect appears at 1.5 PN order and is best represented by the effective spin parameter \cite{Ajith:2009bn},
\begin{equation}
    \chi_{\mathrm{eff}} = \frac{m_1\chi_1+m_2\chi_2}{m_1+m_2} = \frac{\chi_1+q\,\chi_2}{1+q}.
\end{equation}
Here, $(m_1,m_2)$ are the component masses, $(\chi_1,\chi_2)$ are their dimensionless spin components aligned with the orbital angular momentum, and $q = m_2/m_1 \leq 1$ is the mass ratio. Further, the contribution to the GW phase for non-precessing systems to leading order in spin can be written as,
\begin{equation}
\begin{aligned}
    \Psi (f) = \frac{3}{128(\pi f \mathcal{M})^{5/3}} & \Biggl[1 + \nu^2\left(\frac{3715}{756} + \frac{55\eta}{9}\right) \\ 
    & + \nu^3\left(\frac{113}{3} \chi_{\rm eff} - \frac{76\eta}{3}\chi_s -16\pi\right) \Biggl],
\end{aligned}
\end{equation}
where $\mathcal{M} = M\eta^{3/5}$ is the chirp mass, $\eta = q/(1+q)^2$ is the symmetric mass ratio, $\nu = (\pi f M)^{1/3}$ is the PN expansion parameter, $M = m_1 + m_2$ is the total mass, $f$ is the frequency of the quadrupolar part, and $(\chi_s = \chi_1 + \chi_2)/2$.  The dominant effect of spins on the phase can be encapsulated in a reduced spin parameter, $\chi_{\rm PN}$, defined as
\begin{equation}
    \chi_{\mathrm{PN}} = \chi_{\mathrm{eff}} - (76\eta/113)\chi_s.
\end{equation}
As the lowest order contribution to the phase depends on the chirp mass $\mathcal{M}$, it is well measured. Now, looking specifically at the $\nu^2$ (1 PN) and the $\nu^3$ (1.5 PN) terms \cite{Baird:2012cu}, 
\begin{equation}
\begin{aligned}
    \Delta\Psi (f) = \frac{3\nu^2}{128(\pi f \mathcal{M})^{5/3}} & \Biggl[\frac{3715}{756} + \frac{55\eta}{9} \\
    & + \nu\left(\frac{113}{3} \chi_{\mathrm{PN}}-16\pi\right) \Biggl],
\end{aligned}
\end{equation}
we see that $\eta$, or equivalently the $q$, can be varied without varying $\mathcal{M}$ to mimic the effect of spin (though, it will also have to account for the changing $\nu$ with frequency) at this order. This, together with the dependence of $\chi_{\mathrm{eff}}$ on $q$, represents the degeneracy between mass and spin parameters, which hampers the measurement of the individual spins \cite{Cutler:1994ys,Baird:2012cu}.

Secondly, the dimensionless spins of NS are expected to be small by the time of merger. Pulsar observations suggest that NS can be rapidly rotating. While most of the cataloged pulsars have periods $P\sim0.5$ s, a significant and distinct fraction, called millisecond pulsars, can have periods $P\lesssim30$ ms \cite{Lorimer:2008se}. The fastest-rotating known pulsar, PSR J1748-2446ad, has a period of $P\sim1.4$ ms, translating to a dimensionless spin $\chi \leq 0.4$ \cite{Hessels:2006ze,LIGOScientific:2017vwq}. The fastest-spinning NS that is part of a BNS that will merge within Hubble time is PSR J0737-3039, with a period of $22$ ms \cite{Burgay:2003jj}. While this is still rapid, pulsar rotation is expected to slow down with time. This is because pulsars accelerate charged particles which, in turn, emit radiation, and this radiation carries away the rotational kinetic energy. In fact, PSR J0737-3039 is expected to spin-down to $\chi\lesssim0.04$ at merger \cite{LIGOScientific:2017vwq}. As ground-based GW observatories are only expected to detect these binaries close to merger, the NS dimensionless spins are expected to be $\lesssim0.04$.

Fortunately, the mass-spin degeneracy can be broken with the inclusion of higher-order modes and precession \cite{Chatziioannou:2014bma} in the waveform. In this work, we will restrict ourselves to non-precessing systems. While the quadrupolar $(2,2)$ mode dominates the GW from equal-mass and face-on (inclination angle $\iota = 0^{\circ}$) systems, the sub-dominant higher-order modes get activated for systems with unequal masses \cite{Pan:2010hz,Roy:2019phx} (and for precessing systems). We expect the contribution of these higher modes (HM) to increase with the asymmetry in the masses. As the NS mass function is restricted to a small range of values \cite{Suwa:2018uni,2022NatAs...6.1444D,Margalit:2017dij,Ruiz:2017due,Shibata:2017xdx,Rezzolla:2017aly,Shibata:2019ctb,LIGOScientific:2019eut}, we do not except significant contribution of HM in the inspiral of BNS mergers. On the other hand, NSBH mergers will have unequal masses. Population synthesis studies (like Ref.~\cite{Broekgaarden:2021iew}) that have looked at NSBH mergers formed through isolated binary formation channels find that BH mass in NSBH mergers rarely exceeds $20\msun$, with the peak of the mass function lying between $5\msun-15\msun$. This is also supported by the inference of BH mass spectrum from detected NSBH mergers \cite{Biscoveanu:2022iue}. Given that the expected NS masses are less than $2\msun$, NSBH systems will be largely asymmetric with $q\lesssim0.4$. Thus, due to the contribution of HM to the GW signal, NSBH systems can be instrumental in measuring the NS spin ($\chiNS$). 

In this work, we assess the precision with which $\chiNS$\ can be measured from non-precessing NSBH mergers. We find that the precision in the measurement of $\chiNS$ strongly depends on the BH spin $(\chiBH)$, with high BH spins $(\chiBH\geq0.6)$ being conducive to more precise $\chiNS$ measurement. We also show the importance of HM in improving the measurement precision, by comparing the bounds placed on $\chiNS$ using a waveform model that only contains the quadrupolar mode with one that also contains HM, at varying inclination angles. As the detected NSBH events fail to constrain $\chiNS$ \cite{KAGRA:2021vkt,KAGRA:2021duu,LIGOScientific:2021qlt}, we focus on the abilities of next-generation (XG) GW observatories in measuring the same. We find that a five-detector $\mathrm{A}+$  \cite{Miller:2014kma,KAGRA:2013rdx} (three-detector $\mathrm{A}^{\sharp}$) \cite{T2200287} network might distinguish $\chiNS = 0.04$ from zero at $1-\sigma$ confidence for an NSBH merger at $\sim30$ $(\sim50)$ Mpc, while a network with the Einstein Telescope (ET) \cite{Punturo:2010zz,Hild:2010id,Branchesi:2023mws} and two Cosmic Explorers (CE) \cite{LIGOScientific:2016wof,Reitze:2019iox,Evans:2021gyd,Evans:2023euw,Gupta:2023lga} can accomplish the same for an NSBH merger at $\sim350$ Mpc. Based on these results, we claim that \textit{if} there exists a sub-population of NSBH binaries in nature that merge within Hubble time and contain rapidly spinning NS, then XG observatories will discern such NS from slowly or non-spinning ones, identifying extragalactic NS with millisecond rotational periods without electromagnetic observations.

The rest of the paper is structured as follows. In section \ref{sec:motiv}, we build on the motivation for this study. The description of the employed parameter estimation techniques, the NSBH systems that are considered, and the GW network configurations that are used are presented in section \ref{sec:methods}. In section \ref{sec:results}, we show the constraints on $\chiNS$ that can be placed with NSBH mergers, accompanied by the analysis of how BH parameters, the presence of HM, and detector sensitivity can affect the precision with which $\chiNS$ can be estimated. Our conclusions are summarized in section \ref{sec:concl}. 


\section{Motivation} \label{sec:motiv}
In this section, we describe the motivation for considering NSBH mergers to precisely measure the NS spin. In section \ref{subsec:motiv_theory}, we use PN expressions to show the utility of HM in improving the measurement of $\chiNS$. In section \ref{subsec:motiv_astro}, we discuss the astrophysical processes that can lead to the formation of NSBH systems with rapidly spinning NS that merge within Hubble time.

\subsection{Theoretical Arguments} \label{subsec:motiv_theory}
In section \ref{sec:intro}, we explained the degeneracy between the mass and spin parameters using the PN expansion of the GW phase to leading order in spin. We encapsulated the effect of spin to the phase at this order in the term $\chi_{\mathrm{PN}}$, which is a function of $\chi_{\mathrm{eff}}$ and $\chi_{s}$. For BNS systems, the mass ratio is close to $1$, leading to $\chi_{\mathrm{eff}} \sim \chi_{s}$. Thus, while these combinations of component spins can be measured to significant precision, it is difficult to disentangle the values of individual spins from them. These degeneracies could be affected due to higher-order spin effects, including the effect of spin-induced quadrupole moments on the phase \cite{Krishnendu:2017shb,Nagar:2018zoe,Dietrich:2019kaq}. However, given that the values of $\chiNS$ are expected to be small, we do not expect these effects to drastically improve the measurement of NS spins. 

For NSBH systems, we expect $\chiNS$ to be small and the mass ratio to be much smaller than $1$. With these assumptions,
\begin{equation*}
    \chi_{\mathrm{eff}} = \frac{\chiBH + q\,\chiNS}{1+q} \approx \chiBH,
\end{equation*}
i.e., most of the information contained in $\chi_{\mathrm{eff}}$ is about $\chiBH$, as the contribution of $\chiNS$ is weighed down by $q$. While $\chiBH$ and $\chiNS$ are at equal footing in $\chi_{s}$, the contribution of $\chi_{s}$ compared to $\chi_{\mathrm{eff}}$ is weighed down by $\eta$ (as, by definition, $\eta \leq 1/4$).

\begin{figure}[htbp]
{\centering \includegraphics[scale=0.5]{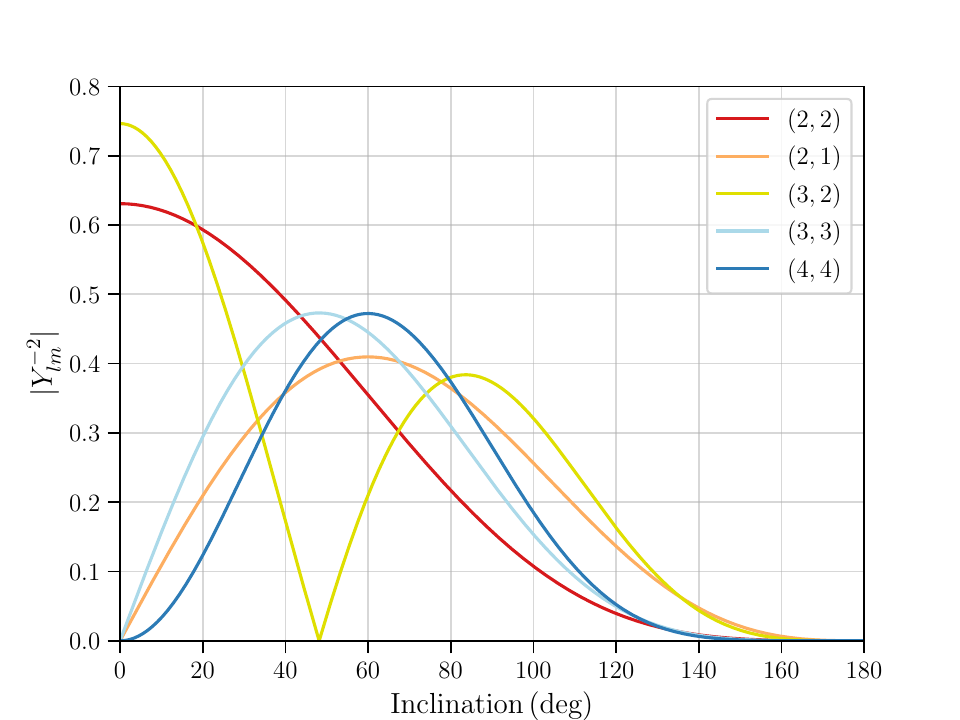}}
\caption{\label{fig:ylm_incl} The variation of the absolute value of $Y^{-2}_{lm}$ with the inclination angle for different HM. $|Y^{-2}_{l-m}|$ are same as $|Y^{-2}_{lm}|$ but mirrored about $90^{\circ}$.}
\end{figure}

\begin{table}[htbp] 
\centering
\caption{\label{tab:hm_amps}The table lists the leading-order spin terms that contribute to the amplitudes of some of the different $(l,m)$ modes, retrieved from \citet{Pan:2010hz}. Here, the column titled PN lists the PN order of the leading order term that contributes to the amplitude of the mode, and the PN$(\chi)$ column lists the leading order at which the spin terms show up in the amplitudes of the respective modes. Similar to $\chi_{\mathrm{eff}}$, we introduce another combination of component spins and mass ratio, $\Tilde{\chi} = \frac{\chi_{\mathrm{BH}}-q\,\chi_{\mathrm{NS}}}{1+q}$, which contains the leading order spin effects to the $(2,1)$ and the $(3,3)$ mode amplitudes.}
\renewcommand{\arraystretch}{1.7}
\begin{tabular}{ P{0.9cm} | P{1cm} | P{1cm} | P{5cm}}
\hhline{----}
Mode & PN & PN$(\chi)$ & Leading order spin terms \\
\hhline{----}
$(2,1)$ & 0.5 & 1 & $\frac{4i}{R}\sqrt{\frac{\pi}{5}}\,\eta M\,\Tilde{\chi}$  \\
\hhline{----}
$(2,2)$ & 0 & 1.5 & $\frac{32}{3R}\sqrt{\frac{\pi}{5}}\,\eta M\,\left(\chi_{\mathrm{eff}}-\eta\chi_s\right)$  \\
\hhline{----}
$(3,2)$ & 1 & 1.5 & $\frac{32}{3R}\sqrt{\frac{\pi}{7}}\,\eta^2 M\,\chi_s$  \\
\hhline{----}
$(3,3)$ & 0.5 & 2 & $\frac{3i}{2R}\sqrt{\frac{6\pi}{7}}\,\eta M\,\left((4-5\eta)\Tilde{\chi} - 14\eta\chi_a \right)$  \\
\hhline{----}
$(4,4)$ & 1 & 2.5 & $\frac{256}{9R}\sqrt{\frac{\pi}{7}}\,\eta M\,\Big[(-\frac{2}{3} + \frac{13}{5}\eta)\chi_{\mathrm{eff}}$ \\
& & & + $\frac{2\eta}{5}\left( \frac{1}{3} - 7\eta \right)\chi_s \Big]$  \\
\hhline{----}
\end{tabular}
\end{table}

\begin{figure*}[htbp]
{\centering \includegraphics[scale=0.6]{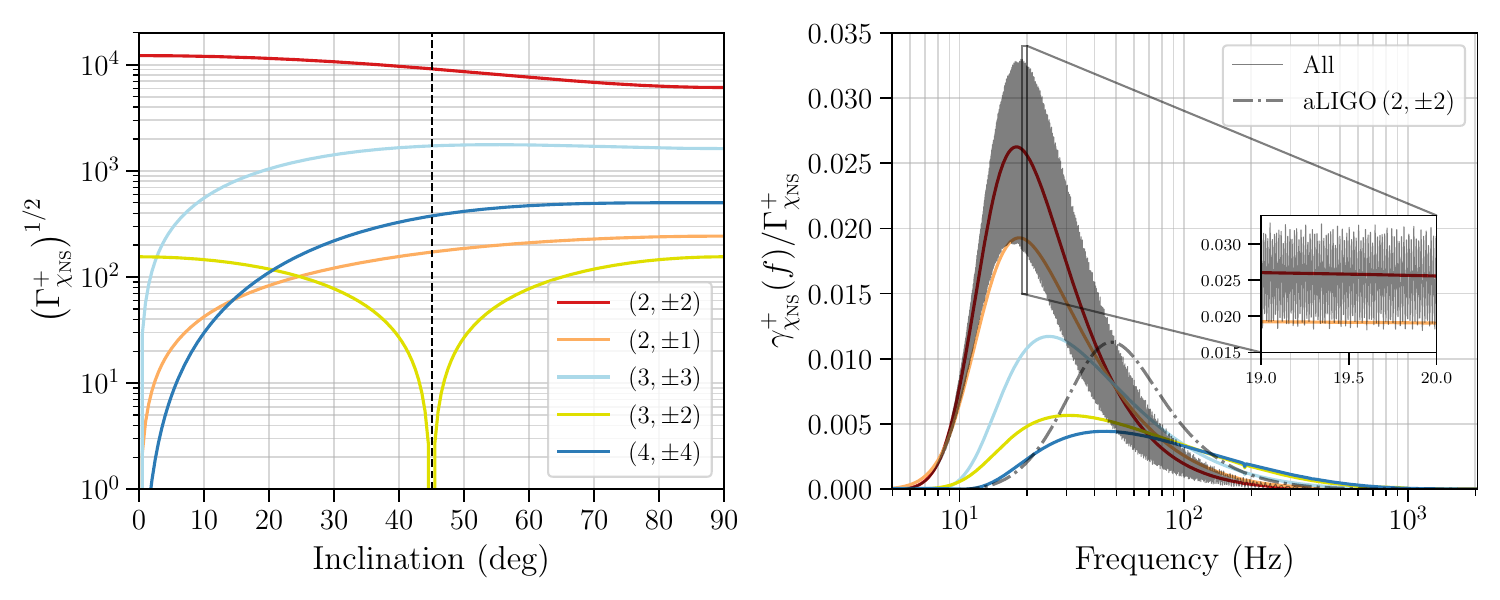}}
\caption{\label{fig:fish} Left panel: The square-root of the Fisher term corresponding to $\chiNS$ for the various $(l,m)$ modes as a function of redshift (see equations \ref{eq:6}-\ref{eq:8} for definitions of the various terms). The dotted vertical line indicates $\iota = 45^{\circ}$. Right panel: Assuming $\iota = 45^{\circ}$, the figure shows the normalized Fisher term as a function of frequency for the different $(l,m)$ modes separately, whereas the black curve shows the normalized Fisher term for the full waveform that contains all the modes. The highly oscillatory nature of the full waveform, due to constructive and destructive interference between modes, is highlighted in the inset. The 40 km CE PSD was used for obtaining the Fisher estimates. Using a different PSD, like that for aLIGO, which has worse sensitivity at lower frequencies, will shift the curves towards higher frequencies, which is shown for the $(2,\pm2)$ mode with the dot-dashed line.}
\end{figure*}

The HM also contain contributions from the spins, which can help improve the measurements of individual spins of binary components. The complex time-domain GW strain $h(t)$ can be decomposed in the basis of $-2$ spin-weighted spherical harmonics $Y^{lm}_{-2}$,
\begin{equation}
    h(t,\boldsymbol{\lambda}) = \sum_{l=2}^{\infty} \sum_{m=-l}^{l} Y_{-2}^{lm} (\iota)\,h_{lm}(t,\boldsymbol{\lambda}),
\end{equation}
where $\iota$ is the inclination angle, and $\boldsymbol{\lambda}$ is the set of parameters, including masses and spins of the binary components, that affect the strain. For non-precessing systems, the $(l,\pm m)$ modes obey the symmetry $h_{lm} = (-1)^{l}\,\Bar{h}_{l-m}$. The phase of $(l,m)$ modes has an approximate relation to the phase of the $(2,2)$ mode in the inspiral regime, given by $\Phi_{lm} \approx \frac{m}{2} \Phi_{22}$, which also means that the frequency corresponding to the $(l,\pm m)$ mode is $m/2$ times the frequency of the $(2,\pm2)$ mode. On the other hand, the amplitudes of the HM contain important differences. Firstly, the relative contribution of the $(l,m)$ mode to the signal depends on the inclination angle, which enters the equation through $Y_{lm}^{-2}$. Figure \ref{fig:ylm_incl} shows the dependence of the absolute value of $Y_{lm}^{-2}$ with inclination for different modes. Secondly, various combinations of component spins contribute to the amplitude of the $(2,2)$ mode and the HM. The leading-order spin contributions to $h_{lm}$ are listed in table \ref{tab:hm_amps}, with lower-order terms only depending on mass parameters. Based on table \ref{tab:hm_amps} and figure \ref{fig:ylm_incl}, we conclude the following:
\begin{itemize}
    \item The leading order term in the $(2,2)$ mode amplitude is at 0 PN. This makes $(2,2)$ the dominant mode in the GW strain.   
    \item Only the $(2,2)$ and $(3,2)$ modes contribute to the signal for face-on $(\iota = 0^{\circ})$ systems. The $(2,1)$, $(3,3)$ and $(4,4)$ modes peak at $\iota > 40^{\circ}$.
    \item The $(2,2)$, $(3,2)$ and $(4,4)$ modes contain positive combinations of spins, like $\chi_{\mathrm{eff}}$ and $\chi_s$, whereas $(2,1)$ and $(3,3)$ modes contain negative combinations, like $\Tilde{\chi}$ and $\chi_a$.
    \item In their respective modes, the $\chi_s$ $(\chi_a)$ spin terms are suppressed by a factor of $\eta$ compared to the $\chi_{\mathrm{eff}}$ $(\Tilde{\chi})$ term. 
\end{itemize}
The different ways in which the mass and spin terms interact in the various $(l,m)$ modes help remove the degeneracy between these parameters. While $\chi_{\rm eff}$ and $\Tilde{\chi}$ values for BNS systems are expected to be small, they can be significant for NSBH systems if the BH is highly spinning. As higher-order spin contributions to amplitude and phase contain higher powers of $\chi_s$ and $\chi_a$, larger BH spin will be conducive to improving the measurement of component spins. Thus, we expect a correlation between greater $\chiBH$ values and better measurement of $\chiNS$. This expectation will be validated in section \ref{subsec:results_bh_mass_spin}.

To better illustrate the effect of HM, we calculate the Fisher matrix element $(\Gamma_{\chiNS})$ for a typical NSBH system using \texttt{gwbench} \cite{Borhanian:2020ypi}. $\Gamma_{\chiNS}$ is a measure of the amount of information about $\chiNS$ that can be extracted from the GW strain. For demonstration, we choose the \textit{plus} polarization of the strain and calculate
\begin{align}
    h^{+}_{lm,\chiNS} (f) &= \frac{\partial h^{+}(f)}{\partial \chiNS}, \label{eq:6}\\ 
     \gamma^{+}_{\chiNS}(f) &= \frac{h^{+}_{lm,\chiNS} (f)\,\bar{h}^{+}_{lm,\chiNS} (f)}{S_n(f)},  \label{eq:7}\\ 
    \Gamma^{+}_{\chiNS} &= 4 \Re \int_{f_{\mathrm{low}}}^{f_\mathrm{high}} \gamma^{+}_{\chiNS} (f)\,df, \label{eq:8}
\end{align}
where $h^{+}_{lm}(f)$ is the \textit{plus} polarization of the $(l,m)$ mode in frequency-domain obtained using IMRPhenomXHM \cite{Garcia-Quiros:2020qpx} for an NSBH system with $10\msun$ BH, $1.4\msun$ NS with the power spectral density $(S_n)$ associated with the 40 km CE detector considered between $f_{\rm low} = 5$ Hz and $f_{\rm high} = 2048$ Hz (see Appendix~\ref{app:det_conf} for more information). We choose $\chiBH = 0.4$ and $\chiNS = 0.04$. Note that the square-root of the inverse of the full Fisher matrix gives the measurement errors on the parameters. However, due to the process of inversion, the measurement precision on $\chiNS$ is not determined just from $(\Gamma^{+}_{\chiNS})^{-1/2}$ but has contributions from the correlations between $\chiNS$ and all the other GW parameters. 

Hence, keeping in mind that $(\Gamma^{+}_{\chiNS})^{1/2}$ is only an indicator of the contribution of different modes, we show how these contributions vary with the inclination angle and the frequency in figure~\ref{fig:fish}. The left panel shows the contributions of the different $(l,m)$ modes to the measurement of $\chiNS$. The effect of the variation of $|Y^{-2}_{lm}|$ (c.f. figure \ref{fig:ylm_incl}) with inclination is apparent. As expected, the bulk of the information about $\chiNS$ comes from the $(2,\pm 2)$ mode across the various inclination angles. The second strongest contribution comes from the $(3,\pm 3)$ mode, which considerably informs the $\chiNS$ estimates for systems with $\iota > 20^{\circ}$. At high inclination angles, the contribution from the $(3,\pm 3)$ mode can reach about $1/3$ the contribution of the $(2,\pm 2)$ mode (see Ref.~\cite{Mills:2020thr} for details on the importance of different modes). In the right panel, we show $\gamma^{+}_{\chiNS}(f)$ as a function of frequency, but normalized by its integral over the frequency band, $\Gamma^{+}_{\chiNS}$. The normalization removes information about the relative importance of each mode, which is already shown in the left panel, and allows the portrayal of frequency regions in which each of the modes dominates. The $(2,\pm 2)$ and $(2,\pm 1)$ modes contribute predominantly to the lower frequencies, whereas the other HM contribute to the higher frequencies as well. The information from the waveform that contains the $(2,\pm 2)$ and the HM, shown in black, follows the $(2,\pm 2)$ mode curve, with oscillations due to the constructive and destructive interference between the $(2,\pm 2)$ mode and the HM \cite{Arun:2008kb}. 

It is important to note that the choice of $S_n (f)$ affects the frequency regions to which the different modes contribute the most. The power spectral density corresponding to CE 40 km is relatively flat at lower frequencies, where the systems spend the majority of their inspiral. However, aLIGO poor low-frequency sensitivity (c.f. figure~\ref{fig:psd}), which will result in the shift of the curves shown in the right panel of figure \ref{fig:fish} towards higher frequency values. This shift has been specifically shown for the $(2, \pm2)$ mode with aLIGO sensitivity in figure \ref{fig:fish}. 

From the theoretical considerations involving the PN theory and Fisher estimates, we conclude that the $(2,2)$ mode will dominate the $\chiNS$ measurement, followed closely by the $(3,3)$ mode, especially for near edge-on $(\iota = 90^{\circ})$ systems. We highlighted that $\eta$ combines differently with different spin contributions in the amplitude of the higher modes. We also posit that a large $\chiBH$ value will be conducive to $\chiNS$ measurement.

\subsection{Astrophysical Considerations} \label{subsec:motiv_astro}
As discussed in section \ref{sec:intro}, pulsar observations have shown that rapidly rotating NS can exist in binary configurations that merge within Hubble time. Further, in section \ref{subsec:motiv_theory}, we assert that for an NSBH system, larger values of BH spin will correspond to better measurements of the NS spin. In this section, we assess the formation scenarios for NSBH systems, via isolated binary formation channels, with a rapidly rotating NS and, preferably, a highly spinning BH. 

One way NS can attain their spins is due to the collapse of an initially spinning Chandrashekhar core. However, as the mass transfer from the core to the outer layers before collapse is expected to be highly efficient \cite{Cantiello:2014uja}, the cores right before collapse are likely to be slowly spinning, leading to slowly spinning compact stars post-collapse \cite{Fuller:2019sxi}. Recent studies \cite{Coleman:2022lwr,Burrows:2023ffl} suggest another method to spin-up the NS at birth--- through stochastic accretion of in-falling matter after the core collapse supernova explosion. In fact, \citet{Burrows:2023ffl} finds that NS created in this way from lower mass $(< 10\msun)$ ZAMS stars have lower spins ($P\sim\mathcal{O}(10^3)$ ms) and natal kick velocities ($v_{\rm kick}\lesssim200\,\mathrm{km\,s}^{-1}$), whereas those from higher mass $(> 10\msun)$ ZAMS stars generally have higher spins ($P\sim\mathcal{O}(10)$ ms) and natal kick velocities ($v_{\rm kick}\gtrsim400\,\mathrm{km\,s}^{-1}$). While the latter is favorable for generating rapidly-spinning NS, the high kick velocities could disrupt the binary system, reducing the chance of such NS to exist in binary configurations. A similar trend is seen in BH, where BH formed after core-collapse are seen to have large spins and kicks, whereas those born from failed explosions have smaller values for spins and kicks \cite{Burrows:2023nlq,Burrows:2023ffl}. 

Stars present in binary configurations can undergo other physical processes that can lead to rapidly rotating compact objects. The primary, more massive, star is the first to evolve off the main sequence and undergo a supernova explosion to become a compact object. Assuming the binary survives this explosion, the second star evolves off the main sequence, expands and initiates the common envelope phase \cite{Soberman:1997mq,2010ApJ...717..724G,2015ApJ...812...40G,Vigna-Gomez:2018dza}. The common envelope extracts energy from the binary and, if ejected successfully, leaves behind a tighter binary with smaller period, containing a compact object and a Wolf-Rayet star (for other scenarios, see Ref. \cite{Broekgaarden:2021iew} and references therein). In certain cases, the Wolf-Rayet star can expand further and fill its Roche Lobe, leading to stable mass-transfer from this helium-star to the compact object \cite{1981A&A....96..142D,Tauris:2015xra}. This mass transfer episode can spin-up the first-born compact object \cite{Thorne:1974ve,Tauris:2015xra}. If the BH is formed first, \citet{Wang:2024pbr} shows that super-Eddington accretion can result in highly spun-up $(\chiBH > 0.6)$ BH. If the NS is formed first, its spin can be further increased (often referred to as recycling the NS/pulsar), leading to more massive NS with millisecond periods \cite{Ozel:2012ax} and possible burial of their magnetic fields \cite{2006MNRAS.366..137Z,Chattopadhyay:2020lff}. As the spin-down rate of the NS increases with the magnetic field strength and decreases with mass, these recycled NS can have considerable spins at the time of merger. In their simulation study, \citet{Chattopadhyay:2020lff} find that $\sim20\%$ of the NSBH systems where NS is formed first contain a pulsar and more than $96\%$ of such \textit{radio-alive} systems are noted to be recycled. They also find that for a significant fraction of such binaries, $\chiNS > 0.05$ during merger. However, only $\sim3\%$ of the NSBH systems in their simulations, that are expected to be detected by the LIGO-Virgo detectors, contain a first-born NS. For the rest of the systems, NS is born second and cannot get spun-up due to accretion. Other studies also estimate the rate of NSBH mergers with first-born NS to be subdominant to those with second-born NS, with the ratio of rates for the former to that of the latter ranging from $\mathcal{O}(0.001)-\mathcal{O}(0.1)$ \cite{Broekgaarden:2021iew,Chattopadhyay:2022cnp}. 

The helium core of the secondary, present in a close binary, can also get spun up due to tidal locking \cite{2017MNRAS.467.2146K,Qin:2018vaa,Fuller:2019sxi,Bavera:2020uch}. \citet{Fuller:2019sxi} find that the spin of second-born BH can exceed the value of 0.5 for such systems. However, as the tidal torques depend on the orbital angular frequency, the binary has to be very close ($P_{\rm orb} \lesssim 1\,{\rm day}$~\cite{Fuller:2022ysb}) for tidal locking to take effect. \citet{Fuller:2022ysb} find that for the orbital period $P_{\rm orb} = 0.5\,{\rm day}$, second-born NS can be spun-up to millisecond periods. As shorter orbital separations are required for tidal synchronization, the companion has to be small enough to fit in the orbit. Compact objects like NS and BH make ideal companions that can tidally spin up the helium core, which is why it is the secondary that is expected to spin up appreciably due to tidal locking. Using \citet{Qin:2018vaa}, \citet{Chattopadhyay:2020lff} find that a significant fraction of second-born BH in NSBH systems can have spins greater than 0.4. They also probe the effect of metallicity on the NS spins in NSBH mergers, finding that lower metallicity results in a lower mass loss through stellar winds. This results in more massive helium stars, which expand less and result in lower mass transfer. Thus, the NS are not efficiently recycled due to accretion for such systems. 

While the problem of binary formation and evolution is dominated by modelling uncertainties, several plausible pathways lead to the formation of NSBH binaries with rapidly rotating components, even under the constraints of the isolated binary formation mechanisms. Although the actual merger rates and properties of such systems can only be ascertained with more observations, it is worth investigating if these properties can be precisely determined with future GW observatories. Given the specific and countable scenarios that lead to the presence of a rapidly rotating NS during merger, decisively differentiating the spin of the NS from $0$ can illuminate the astrophysical processes that contribute to cosmic binary formation and evolution, and the physics that affects the spin-down of NS. This is the objective of this work, and the particular question of the precision in the NS spin measurement will be answered in section \ref{sec:results}.


\section{Methodology} \label{sec:methods}
To explore the precision with which $\chiNS$ can be measured using NSBH mergers, we apply Bayesian parameter estimation on simulated GW data. The GW data $d$ is the detector response defined by,
\begin{equation}
    d(t,\boldsymbol{\theta}) = n(t) + s(t,\boldsymbol{\theta}),
\end{equation}
where $n(t)$ is the noise and $s(t,\boldsymbol{\theta})$ is the GW signal which depends on a set of intrinsic and extrinsic parameters (including masses and spins), $\boldsymbol{\theta}$. Then, the posterior distribution $p(\boldsymbol{\theta}|d)$ can be obtained using
\begin{equation}
    p(\boldsymbol{\theta}|d) = \frac{p(d|\boldsymbol{\theta})\,p(\boldsymbol{\theta})}{\mathcal{Z}}.
\end{equation}
Here, $p(d|\boldsymbol{\theta})$ is the likelihood function, $p(\boldsymbol{\theta})$ is the prior on the GW parameters, and $\mathcal Z = \int p(d|\boldsymbol{\theta})\,p(\boldsymbol{\theta})\,d\boldsymbol{\theta}$ is the evidence. The probability distribution for $\chiNS$ can be retrieved from the joint posterior distribution by marginalizing it over all other parameters \cite{2019PASA...36...10T},
\begin{equation}
    p(\chiNS|d) = \int \left( \prod \limits_{\theta_i \neq \chiNS} d\theta_i\right)\,p(\boldsymbol{\theta}|d).
\end{equation}
The calculation of these posterior distributions is a computationally challenging task and is accomplished using stochastic samplers \cite{Metropolis:1953am,b1878965-730b-33c7-b1ad-dfd3acb6f61b,Skilling:2004pqw}. However, covering the prior space and sufficiently mapping the likelihood surface could involve $\mathcal{O}(10^8)$ likelihood evaluation, which can be time-consuming. This is especially true for the long and high signal-to-noise ratio (SNR) signals expected to be detected by XG observatories. Hence, several likelihood evaluation and sampling techniques have been introduced to speed up this process \cite{Canizares:2014fya,Smith:2016qas,Morisaki:2020oqk,Morisaki:2023kuq,Vinciguerra:2017ngf,Morisaki:2021ngj,Pathak:2022iar,Pathak:2023ixb,Pankow:2015cra,Smith:2019ucc,Talbot:2019okv,Tiwari:2023mzf,Williams:2021qyt,Green:2020hst,Edwards:2023sak,Wong:2023lgb}. 

For accelerated parameter estimation, we use the relative binning technique \cite{Cornish:2010kf,Zackay:2018qdy,Cornish:2021lje} implemented in Bilby \cite{2019ApJS..241...27A,2020MNRAS.499.3295R,Krishna:2023bug}. The relative binning technique relies on the assumption that the ratio of the waveform between neighboring points in the parameter space varies smoothly. The waveform corresponding to the maximum likelihood point in the parameter space is chosen as the fiducial waveform, and the waveforms around this point are obtained using piecewise linear functions informed by this fiducial waveform. These approximations can amount to speed ups by $\mathcal{O}(10^3)$, without compromising significantly in the accuracy \cite{Zackay:2018qdy,Krishna:2023bug}. However, waveforms that contain contributions of HM and involve precession can be highly oscillatory and may not be well-approximated by piecewise linear functions. While \citet{Krishna:2023bug} show that the technique fairs well even for asymmetric systems, a mode-by-mode adaptation of relative binning has also been proposed \cite{Leslie:2021ssu}, which approximates each $(l,m)$ mode separately and then adds them together to obtain the full (approximated) waveform. As our study only considers NSBH systems whose spins are aligned with the orbital angular momentum, we use the original relative binning technique and its implementation in Bilby~\cite{Krishna:2023bug}, but also validate our results against those obtained with our implementation of the mode-by-mode technique, and find them to be consistent.

\begin{figure*}[htbp]
{\centering \includegraphics[scale=0.7]{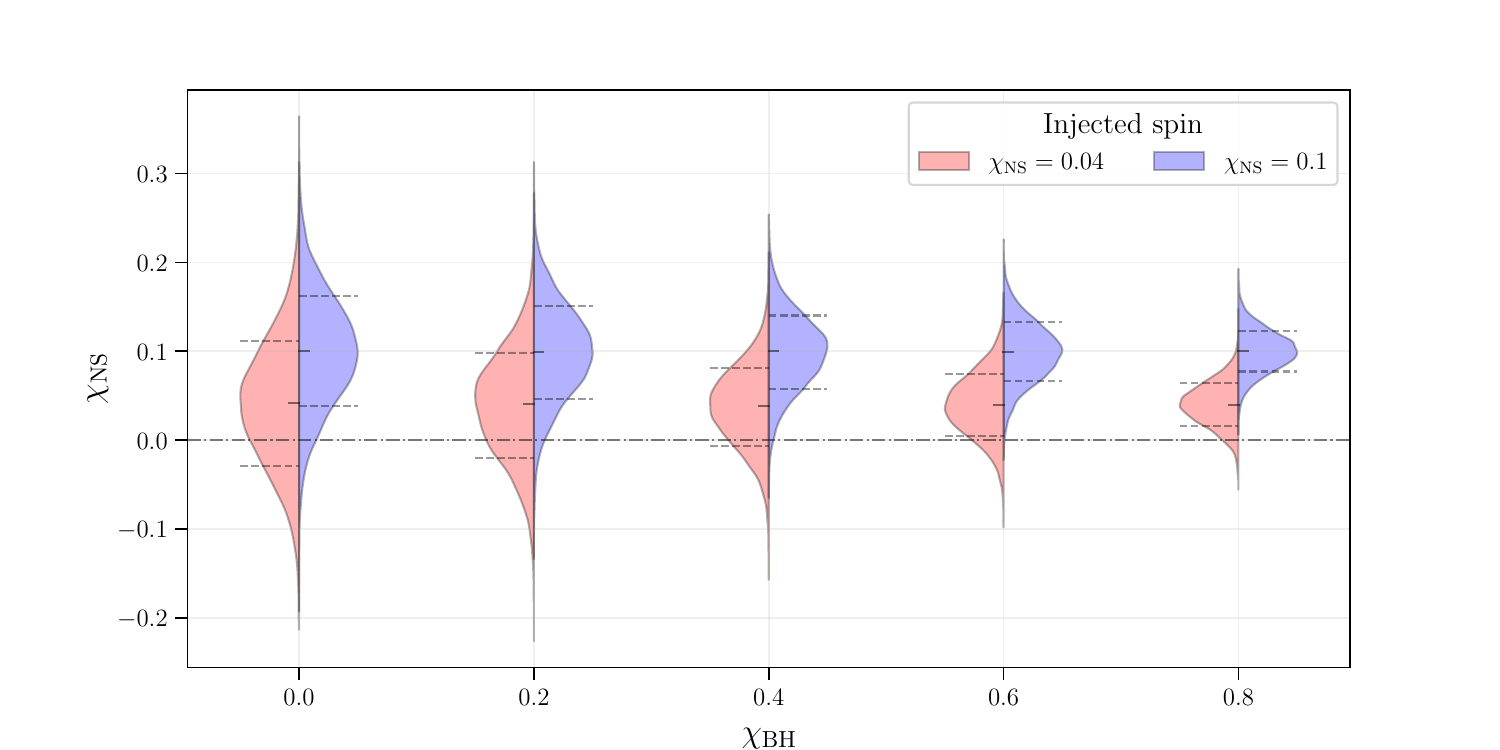}}
\caption{\label{fig:eff_bbh_spin} The posterior distributions for $\chiNS$ as a function of $\chiBH$. The red and blue plots show the posteriors for systems where $\chiNS$ is set to $0.04$ and $0.1$, respectively. Along with the posteriors, we also show the median (small solid horizontal line) and the boundaries of the $1-\sigma$ region (dashed horizontal lines). The posteriors correspond to a system with BH of mass $10\msun$ and NS of mass $1.4\msun$ at an inclination angle of $45^{\circ}$. The dash-dotted horizontal line indicates $\chiNS = 0$.}
\end{figure*}

The GW signals are generated using the IMRPhenomXHM waveform \cite{Garcia-Quiros:2020qpx}, which is an aligned-spin frequency-domain waveform that includes the $(2,2)$, $(2,1)$, $(3,3)$, $(3,2)$ and $(4,4)$ modes. We set the right ascension $\alpha = 78.78^{\circ}$, declination $\delta = -69.37^{\circ}$, phase $\phi_c = 74.48^{\circ}$, polarization angle $\psi = 152.34^{\circ}$, and the geocentric time $t_c = 1126259642.413$ s. Other than section \ref{subsec:results_det_sens} where we look at various detector sensitivities, we inject GW signals in a network of one ET and two CE (one with 40 km arms and the other with 20 km arms), henceforth referred to as the ECC network. Our simulated binaries comprise of NSBH systems with a $10\msun$ BH and a $1.4\msun$ NS (henceforth, referred to as 10+1.4 system), at $\iota = 45^{\circ}$. We assess the effect of varying BH mass in section \ref{subsec:results_bh_mass_spin} and $\iota$ in section \ref{subsec:results_incl_hom}. The BH spins can range from [0,0.8], whereas the NS spins are only considered up to 0.1 (an informed upper-bound based on the results of Ref.~\cite{Chattopadhyay:2020lff}).  

For all the simulations, we use uniform priors in $\chi_1$ and $\chi_2$ (i.e. $\chiBH$ and $\chiNS$, respectively) bounded between [-0.99,0.99], and uniform priors in component masses. For all other parameters, we choose the default Bilby priors. We predominantly use the \texttt{nessai} \cite{nessai,Williams:2021qyt,Williams:2023ppp} and \texttt{pymultinest} \cite{2014A&A...564A.125B} samplers for our runs, and perform spot-checks with the \texttt{dynesty} \cite{2020MNRAS.493.3132S,sergey_koposov_2023_8408702} sampler to ensure consistency.

Note that the NS may get tidally disrupted by the BH before merger \cite{Kyutoku:2021icp}. These tidal effects will affect the waveform and can potentially contribute to $\chiNS$ constraints due to spin-induced quadrupole moments. These effects are not considered in IMRPhenomXHM as it is predominantly used for BBH systems. While high BH spins favor the disruption of the NS, low mass ratio (i.e. a heavy BH) will disfavor it. We explore the effect of the tidal deformability of the NS on $\chiNS$ measurement in Appendix~\ref{app:tidal}. 


\section{Results} \label{sec:results}
In this section, we show the results of the parameter estimation runs performed on simulated NSBH GW events. Focusing on the precision with which $\chiNS$ can be measured, we assess the impact of BH mass and spin as well as that of $\iota$ and HM on these measurements in sections \ref{subsec:results_bh_mass_spin} and \ref{subsec:results_incl_hom}, respectively. Note that for a system situated at a particular luminosity distance $D_L$, changing the masses/spins of the compact objects or the inclination angle will affect the SNR associated with the detected signal, which has direct consequences for the bounds that can be placed on $\chiNS$. To isolate the effect of binary parameters from the effect of SNR, all the systems considered in sections \ref{subsec:results_bh_mass_spin} and \ref{subsec:results_incl_hom} have the same SNR $(\sim 700)$, which is achieved by varying $D_L$ appropriately. The effect of detector sensitivity, luminosity distance (and, consequently, the SNR) are discussed in section \ref{subsec:results_det_sens}.

\subsection{Effect of BH mass and BH spin} \label{subsec:results_bh_mass_spin}
\begin{figure*}[htbp]
{\centering \includegraphics[scale=0.6]{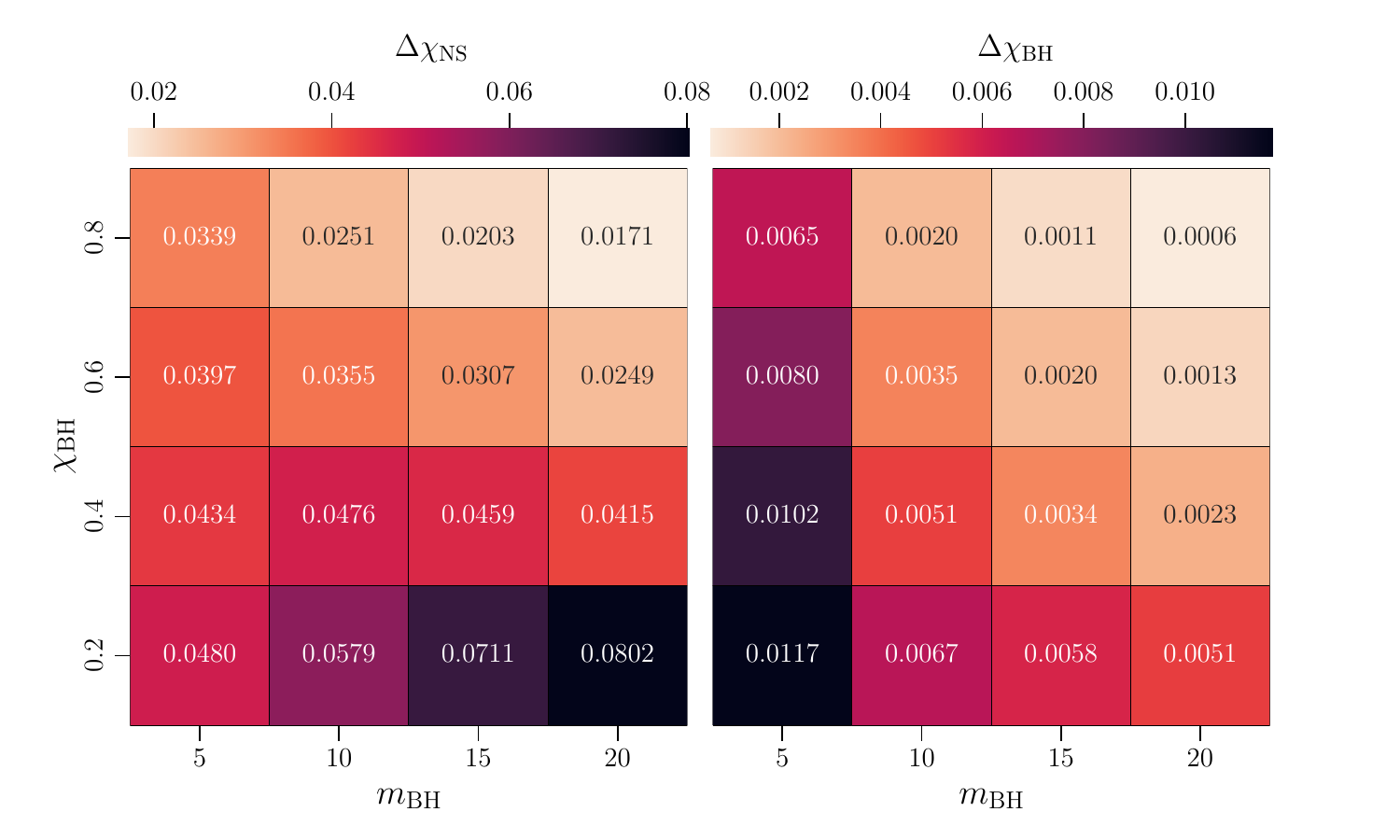}}
\caption{\label{fig:mBH_chiBH_spin_errs} The half-widths of the $68\%$ confidence intervals (i.e., one-sided $1-\sigma$ widths) corresponding to the bounds on $\chiNS$ (left panel) and $\chiBH$ (right panel) for the sixteen NSBH systems with $m_{\rm BH} \in \{5\msun,10\msun,15\msun,20\msun\}$ and $\chiBH \in \{0.2,0.4,0.6,0.8\}$. The NS mass for all the systems was fixed to $1.4\msun$, the NS spin to $0.04$, and $\iota$ to $45^{\circ}$.}
\end{figure*}

In section \ref{subsec:motiv_theory}, we claimed that a high value of $\chiBH$ should improve the $\chiNS$ measurement. To validate this, we consider 10+1.4 NSBH systems at $\iota = 45^{\circ}$ and varying $(\chiBH,\chiNS)$ values, and report the $68\%$ confidence interval (i.e., $1-\sigma$) on the inferred values of $\chiNS$ in figure \ref{fig:eff_bbh_spin}. It is evident that the bounds on $\chiNS$ improve with higher $\chiBH$ values, which validates our expectations. For the chosen SNR (corresponding to systems located at $D_L\in(200,300)$ Mpc), a highly-spinning NS with $\chiNS = 0.1$ can be easily differentiated from a non-spinning one, even for a non-spinning BH companion. A slightly slower NS with $\chiNS = 0.04$ can only be differentiated from $0$ with $1-\sigma$ confidence at this SNR if $\chiBH > 0.4$. As the $\chiNS$ values are small, the bounds are not significantly affected by the true value of the NS spin. 

The effect of the mass of the BH $(m_{\rm BH})$ needs greater care. Keeping the SNR constant, as $m_{\rm BH}$ increases, so does the contribution of the HM to the signal \cite{Mills:2020thr}, which should be favorable for spin measurements. However, this also leads to a lower value for $q$ (and $\eta$), which reduces the contribution of $\chiNS$ to $\chi_{\rm eff}$ and $\Tilde{\chi}$, and of $\chi_s$ and $\chi_a$ to the amplitude of the modes (c.f. table \ref{tab:hm_amps}). To evaluate the effect of $m_{\rm BH}$, we fix the mass and spin of the NS to $1.4\msun$ and $0.04$, respectively, and perform parameter estimation for systems with $m_{\rm BH} \in \{5\msun,10\msun,15\msun,20\msun \}$ and $\chiBH \in \{0.2,0.4,0.6,0.8 \}$. For this set of 16 events, we state the half-width of the $68\%$ confidence interval of the bounds on $\chiNS$ and $\chiBH$ (referred to as $\Delta\chiNS$ and $\Delta\chiBH$, respectively) in figure~\ref{fig:mBH_chiBH_spin_errs}.

The measurement precision of $\chiBH$ is about an order of magnitude better than $\chiNS$. As discussed in section \ref{subsec:motiv_theory}, most of the information to $\chi_{\rm eff}$ comes from $\chiBH$ for highly asymmetric NSBH systems, and $\chi_{\rm eff}$ being the best-measured spin parameter leads to a precise measurement of $\chiBH$ as well. Further, $\chiBH$ measurement improves both with increasing values of $\chiBH$ and $m_{\rm BH}$, which was anticipated earlier in this section. The behavior of bounds on $\chiNS$ is more interesting--- when $\chiBH$ is small, bounds on $\chiNS$ worsen with increasing $m_{\rm BH}$, whereas when $\chiBH \geq 0.6$, the bounds on $\chiNS$ improve with higher $m_{\rm BH}$. This is due to two competing effects--- as $m_{\rm BH}$ increases, $\chiBH$ measurement improves (and so, $\chiNS$ measurement should also improve), whereas the contribution of $\chiNS$ to $\chi_{\rm eff}$ and $\Tilde{\chi}$ decreases (worsening the measurement of $\chiNS$). It turns out that at low $\chiBH$, the latter effect dominates and $\chiNS$ worsens as $m_{\rm BH}$ increases, whereas at high $\chiBH$, the former effect dominates and $\chiNS$ measurement improves with $m_{\rm BH}$. Stated differently, for high $\chiBH$ values, it is extremely well measured, and it is possible to distinguish between $\chiBH$ and the contribution from $q\,\chiNS$ in $\chi_{\rm eff}$ and $\Tilde{\chi}$.

Hence, we see that the $\chiNS$ measurement strongly depends on the values of $\chiBH$. In agreement with our expectations from theoretical arguments, higher $\chiBH$ values are conducive to better $\chiNS$ measurement. For NSBH systems with slowly spinning $(\chiBH < 0.4)$ BH, the bounds on $\chiNS$ worsen with increasing $m_{\rm BH}$, whereas for NSBH systems with highly spinning BH, higher $m_{\rm BH}$ improves the measurement precision for $\chiNS$.

\subsection{Effect of inclination and higher-order modes} \label{subsec:results_incl_hom}
\begin{figure*}[htbp]
{\centering \includegraphics[scale=0.7]{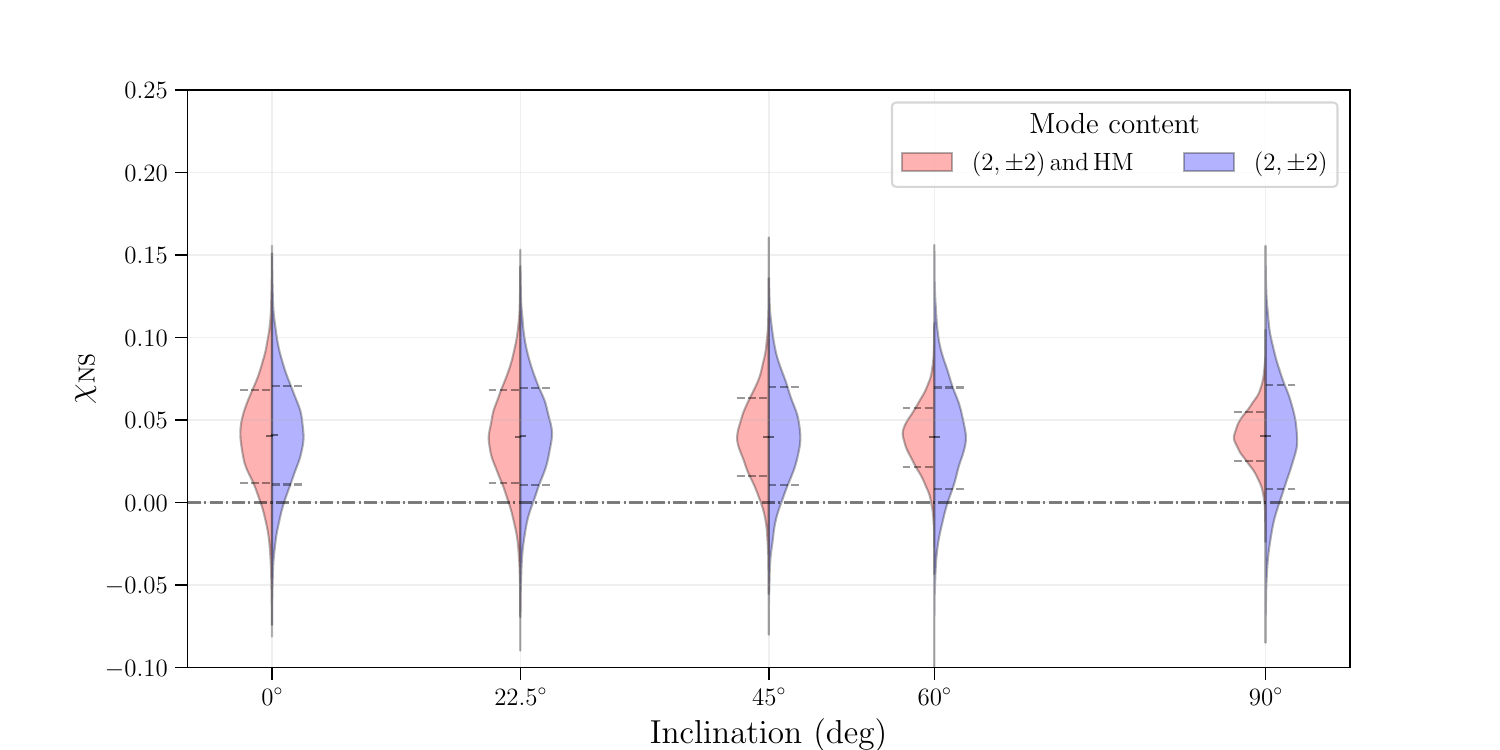}}
\caption{\label{fig:eff_incl_hm} The posterior distributions for the spin of the NS as a function of the inclination angle. The red plots represent the systems for which the injection and the recovery were performed with a waveform that contains both the $(2,\pm2)$ mode and the higher-order modes (IMRPhenomXHM), whereas the blue plots correspond to the systems where the waveform only has the $(2,\pm2)$ mode (IMRPhenomXAS). The dash-dotted horizontal line indicates $\chiNS = 0$.}
\end{figure*}
The relative contribution of HM to the GW waveform increases with inclination angle, which is expected to improve the constraints that can be imposed on $\chiNS$. To evaluate the usefulness of HM towards $\chiNS$ measurement, we compare the bounds on $\chiNS$ obtained when using a waveform model that contains the dominant $(2,\pm 2)$ mode as well as HM to a waveform model that only contains the $(2,\pm2)$ mode. For the former, we continue the use of IMRPhenomXHM, and for the latter, we use IMRPhenomXAS \cite{Pratten:2020fqn}, which is an aligned-spin waveform model which contains only the quadrupole mode. 

We consider the 10+1.4 NSBH systems with $(\chiBH,\chiNS) = (0.8,0.04)$. These are simulated at inclination angles $\iota \in \{0^{\circ},22.5^{\circ},45^{\circ},60^{\circ},90^{\circ}\}$ and the $D_L$ are set such that all the systems are detected with the same SNR. The constraints on $\chiNS$ for these systems when injection and parameter estimation are performed with and without HM are shown in figure~\ref{fig:eff_incl_hm}. The red plots show the bounds on $\chiNS$ for systems where the waveform contains HM, whereas the blue plots correspond to those that only contain the quadrupole mode. It is evident from the figure that $\chiNS$ measurement is significantly improved with the increase in the HM contribution to the waveform at greater $\iota$. On the other hand, the constraints on $\chiNS$ for systems where only the quadrupole mode is used are not affected as $\iota$ is varied, as the signal SNR is kept constant. It is also important to note that the $\chiNS$ posteriors of the two sets of injections do not differ considerably till $\iota = 22.5^{\circ}$. From this, we infer that for such 10+1.4 systems, HM are expected to noticeably affect $\chiNS$ measurements only at inclination angles $\iota \gtrsim 45^{\circ}$. However, GW detections are biased against systems with high $\iota$ values; an edge-on system has similar SNR to a face-on system that is much farther away. Thus, while more GW systems are expected to be oriented face-off, they will be detected with lower SNR compared to face-on systems at the same $D_L$. 

\subsection{Effect of detector sensitivity} \label{subsec:results_det_sens}
\begin{figure}[htbp]
{\centering \includegraphics[scale=0.55]{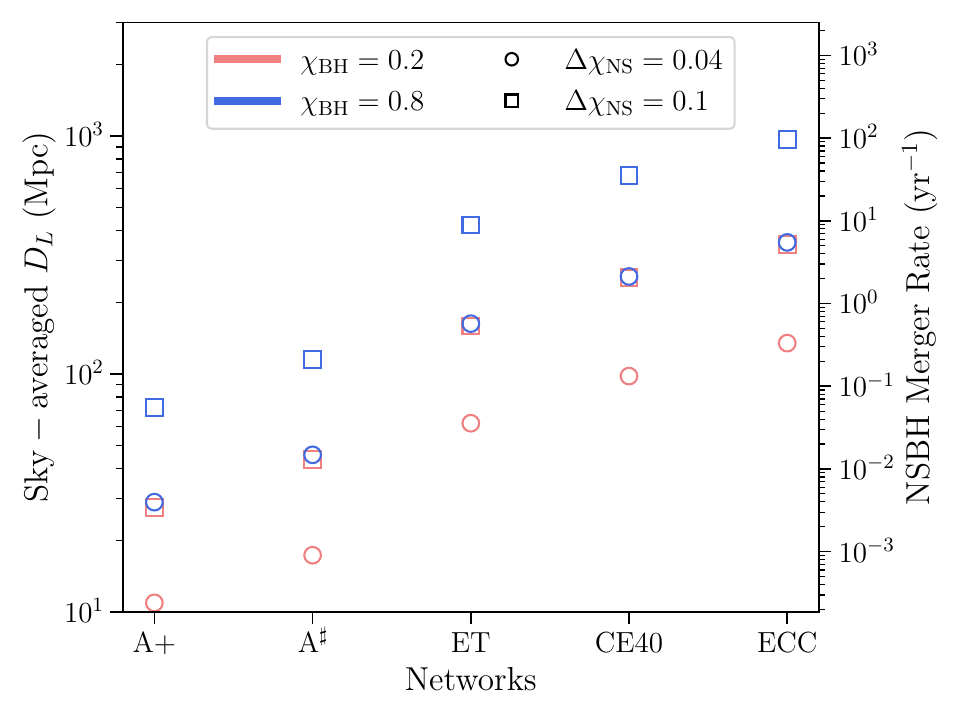}}
\caption{\label{fig:DL} The sky-averaged $D_L$ corresponding to each XG detector network for which a given $\Delta\chiNS \in \{0.04,0.1\}$ bound can be achieved with a 10+1.4 NSBH system at $\iota = 45^{\circ}$ and a given $\chiBH \in \{0.2,0.8\}$. We also denote the corresponding NSBH merger rate per year that is expected at those distances.}
\end{figure}
Till now, we have probed the effect of BH properties and HM on the measurability of $\chiNS$ by keeping the systems at the same SNR $(\sim 700)$. In this section, we look at the measurability as a function of $D_L$ and detector sensitivities. For this purpose, we consider five XG networks- A+ (a network of five detectors at A+ sensitivity), A$^{\sharp}$ (a network of three detectors with A$^{\sharp}$ sensitivity), ET (the triangular ET observatory), CE40 (the 40 km CE observatory), and ECC (a network with the triangular ET, the 40 km CE and the 20 km CE observatories). For a more detailed description of the networks and the different sensitivities, refer to Appendix \ref{app:det_conf}.  We consider 10+1.4 NSBH systems at $\iota = 45^{\circ}$. Parameter estimation is performed for two types of systems--- first, with a slowly-spinning BH $(\chiBH = 0.2)$, and second, with a rapidly-spinning BH $(\chiBH = 0.8)$. In figure \ref{fig:DL}, we show the $D_L$ (averaged over the sky) at which the NSBH mergers occur for different XG observatories to obtain (one-sided $1-\sigma$) $\Delta \chiNS = 0.04$ and $0.1$. Along with the $D_L$, we also report the corresponding NSBH merger rate up to that distance, which is assumed to follow the Madau-Dickinson star-formation rate ~\cite{Madau:2014bja,2019PhRvD.100d3030T} and the local merger rate density for NSBH systems is chosen to be $45\,\mathrm{Gpc}^{-3}\,\mathrm{yr}^{-1}$ \cite{KAGRA:2021duu,LIGOScientific:2021qlt} (see Refs.~\cite{Gupta:2023evt,Gupta:2023lga} for more details). 

For $\chiBH = 0.2$, figure~\ref{fig:DL} shows that the chances for A+ or A$^{\sharp}$ to constrain $\chiNS$ to even $\Delta\chiNS =  0.1$ are bleak, and it would only be possible to do so only for a \textit{golden} event that may occur once every ten to hundred years. However, networks containing ET and CE will be able to constrain $\chiNS$ to $\Delta\chiNS =  0.04$ for an event every few years, and $\Delta\chiNS =  0.1$ for a few events every year. For NSBH systems with rapidly spinning BH, A+ and A$^{\sharp}$ are expected to constrain $\chiNS$ to $\Delta\chiNS = 0.1$ for events occurring once in $10$ years. Networks with just the ET and/or CE will be able to constrain $\chiNS$ to $\Delta\chiNS =  0.04$ for a few events every year, and to $\Delta\chiNS =  0.1$ for tens of events every year. In fact, the ECC detector can constrain $\chiNS$ to $\Delta\chiNS = 0.1$ for events merging at $\sim1000$ Mpc. Up to this distance, we expect $\sim100$ NSBH mergers every year, which increases the chances of detecting a highly spinning NS.

These estimates will depend on the choice of masses for the system as well as the inclination angle. Both these effects have been explored in the previous sections.
Based on the current GW observations, $\chiBH$ in NSBH systems is expected to be low~\cite{Biscoveanu:2022iue}. If this is indeed true for the cosmic NSBH population, then resolving NS spin well enough to differentiate it from $0$ may only be possible, realistically, for networks with CE and ET. However, if rapidly spinning BH were to merge with rapidly spinning NS, A+ and A$^{\sharp}$ networks may constrain $\chiNS$ well enough to distinguish such an NS from a slowly spinning one.


\section{Conclusions} \label{sec:concl}
GW observation can be used to infer source parameters that can illuminate the astrophysical processes involved in binary formation and evolution. Precise measurements of these parameters, particularly the mass and the spin of the binary components, can be used to distinguish between the different astrophysical formation channels. In this work, we have assessed the measurability of the spins of the NS using NSBH detections with XG GW detector networks. 

In section~\ref{subsec:motiv_theory}, we showed that HM can contribute significantly to the waveform and the measurement of the NS spin, especially at higher inclination angles, with $(3,\pm 3)$ being the second highest contributor to the $\chiNS$ measurement. We also noted that $\eta$ interacts with negative combinations of the component spins in the amplitude of the $(2,\pm 1)$ and the $(3,\pm 3)$ modes, and with the positive combinations of component spins in the $(2,\pm 2)$, $(3,\pm 2)$ and $(4,\pm 4)$ modes. These different interactions help mitigate the mass-spin degeneracy. However, due to the narrow mass function of the NS, we do not expect an appreciable contribution of HM to the BNS signal. Hence, we instead look at the bounds that can be placed on $\chiNS$ with NSBH systems, which are naturally mass-asymmetric. 

In section~\ref{subsec:motiv_theory}, we also posit that a higher spin value of the primary will improve the spin measurement of the secondary. This is later confirmed with parameter estimation studies (c.f. figure~\ref{fig:eff_bbh_spin}). In the context of NSBH mergers, this would mean having both the BH and the NS to be rapidly spinning. In section~\ref{subsec:motiv_astro}, we discuss the various astrophysical processes, under the constraints of the isolated binary formation channels, that can lead to the formation of such systems. These include compact objects born after core-collapse supernovae that may rapidly rotate due to stochastic accretion of in-falling matter, spin-up of the first-born compact object by mass accretion due to ultra-stripping of the companion Wolf-Rayet star, and increase in the spin of the secondary due to tidal synchronization with its companion in a close orbit. While the aim of this work is not to justify the existence of such systems, we argue that their detection will provide key insights into the astrophysical processes that govern the formation and evolution of such systems. Thus, in this work, we assessed the possibility of differentiating a rapidly rotating NS from a slow or non-rotating one, using XG GW observatories. 

Using the relative binning technique of accelerated parameter estimation, we performed Bayesian analyses on a set of NSBH systems, the results of which are detailed in section \ref{sec:results}. We find that the constraints on $\chiNS$ are highly contingent on the spin of the companion BH, with high $\chiBH$ values improving the bounds on $\chiNS$ (c.f. figure ~\ref{fig:eff_bbh_spin}). We also discussed the effect of $\chiBH$ on the measurement of $\chiNS$ when varying BH mass--- for low $\chiBH$ $(\sim 0.2)$, the bounds on $\chiNS$ get worse as we increased the BH mass, whereas for a rapidly rotating BH $(\chiBH \gtrsim 0.6)$, $\chiNS$ is measured better for a more massive BH (c.f. figure~\ref{fig:mBH_chiBH_spin_errs}). 

To show the improvements in $\chiNS$ measurement brought out by HM, we compare the posterior distribution of $\chiNS$ for 10+1.4 $\msun$ NSBH systems for two sets of injections at varying inclination angles--- first, with a waveform that contains only the $(2,\pm2)$ mode (IMRPhenomXAS), and second, with a waveform that contains the $(2,\pm2)$ and HM (IMRPhenomXHM).  We find that as the inclination angle increases (and so does the relative HM contribution (c.f. figure ~\ref{fig:fish})), $\chiNS$ measurements improve considerably when HM are included in the waveform, whereas no significant change is seen in the $\chiNS$ measurement when HM is not included (c.f. figure~\ref{fig:eff_incl_hm}), showing the utility of HM in $\chiNS$ measurement. Finally, we explored the effect of SNR and detector sensitivity on the measurability of $\chiNS$. We found that $\chiNS=0.1$ can be distinguished from $0$ with A+ and A$^{\sharp}$ for golden events that may occur once in every tens of years. However, if the BH is rapidly rotating $(\chiBH = 0.8)$, networks with ET and CE will be able to make the same distinction for $\mathcal{O}(100)$ events, merging up to $1000$ Mpc (c.f. figure~\ref{fig:eff_incl_hm}).

Thus, if rapidly spinning NS were to merge with BH, XG observatories with ET and CE would be capable of differentiating such NS from non-spinning ones, making it the first discovery of such NS without electromagnetic observations. Such a discovery would be monumental for astrophysical studies on binary formation and evolution.
As high values of aligned spins are conducive to the tidal deformation of NS before merger, such systems can also be important for multimessenger astronomy \cite{Gupta:2023evt} and cosmology \cite{Gupta:2022fwd}. In this work, we have restricted ourselves to aligned-spin systems. However, with better parameter estimation techniques, this work can be extended to precessing (and even eccentric) NSBH mergers formed due to dynamical interactions. While precession will improve the relative contribution of HM, the additional spin and tilt parameters may increase the uncertainty in $\chiNS$ measurement. Further, the asymmetry between $\pm m$ modes for precessing system can also play a role in measuring the spins of the components \cite{Kalaghatgi:2020gsq,Kolitsidou:2024vub}. Thus, a similar study spanning various binary configurations (in terms of spin and tilt values) will be useful to ascertain if $\chiNS$ can be even better measured than what has been claimed in this work. Another possible extension would be to look at the prospects of detecting a sub-population of such rapidly spinning NS with NSBH systems. This may require less stringent bounds on $\chiNS$ for each event, which will increase the number of potential candidates that can contribute to the analysis.


\begin{acknowledgments}
I would like to thank Bangalore Sathyaprakash, Debatri Chattopadhyay and Aditya Vijaykumar for the helpful discussions and important insights, Koustav Chandra for the comments that have helped improve the manuscript, and Rahul Kashyap, Arnab Dhani, KG Arun, Mukesh Kumar Singh, Rossella Gamba and Sebastiano Bernuzzi for technical suggestions that have been incorporated in this work. This work is funded by the following NSF grants: PHY-2207638, AST-2307147, PHY-2308886, PHY-2309064.
\end{acknowledgments}


\appendix
\section{Detector configurations} \label{app:det_conf}
\begin{figure}[htbp]
{\centering \includegraphics[scale=0.55]{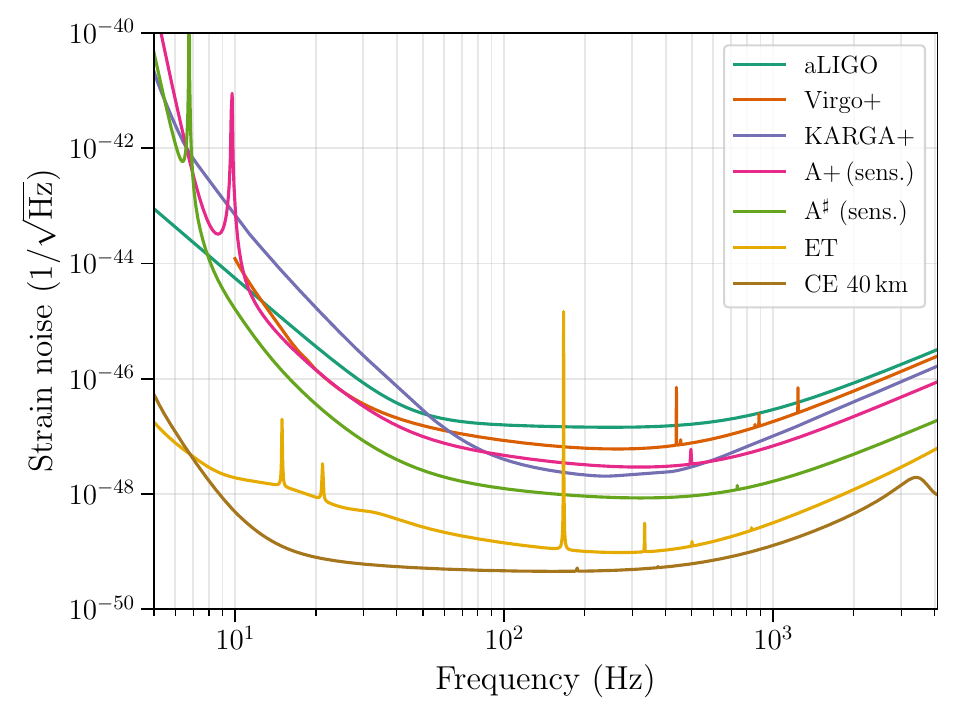}}
\caption{\label{fig:psd} The PSD for the different sensitivities considered in this work. Here, A+ and A$^{\sharp}$ refer to the sensitivity of a given LIGO detector. The files used to generate these curves can be found in the \href{https://gitlab.com/sborhanian/gwbench/-/tree/master/gwbench/noise_curves}{\texttt{gwbench}} repository.}
\end{figure}

To assess the effect of detector sensitivity on the measurability of $\chiNS$, we considered five XG detector networks, which are listed in table~\ref{tab:dets}. The corresponding sensitivities have been plotted in figure~\ref{fig:psd}.
\begin{table}[htbp] 
\centering
\caption{\label{tab:dets}The detector configurations used in section~\ref{subsec:results_det_sens}.}
\renewcommand{\arraystretch}{1.7}
\begin{tabular}{ P{0.9cm} | P{7cm} }
\hhline{--}
Name & Detectors \\
\hhline{--}
A+ & Hanford (US), Livingston (US) and Aundha (India) at A+ sensitivity, Virgo (Italy) at V+ sensitivity, and KAGRA (Japan) at K+ sensitivity  \\
\hhline{--}
A$^{\sharp}$ & Hanford (US), Livingston (US) and Aundha (India) at A$^{\sharp}$ sensitivity \\
\hhline{--}
ET & Triangular ET observatory with three detectors in xylophone arrangement \\
\hhline{--}
CE40 & 40 km L-shaped CE detector \\
\hhline{--}
ECC & Triangular ET, L-shaped 20 km CE and L-shaped 40 km CE\\
\hhline{--}
\end{tabular}
\end{table}


\section{Effect of tides on NS spin measurement} \label{app:tidal}
\begin{figure}[htbp]
{\centering \includegraphics[scale=0.5]{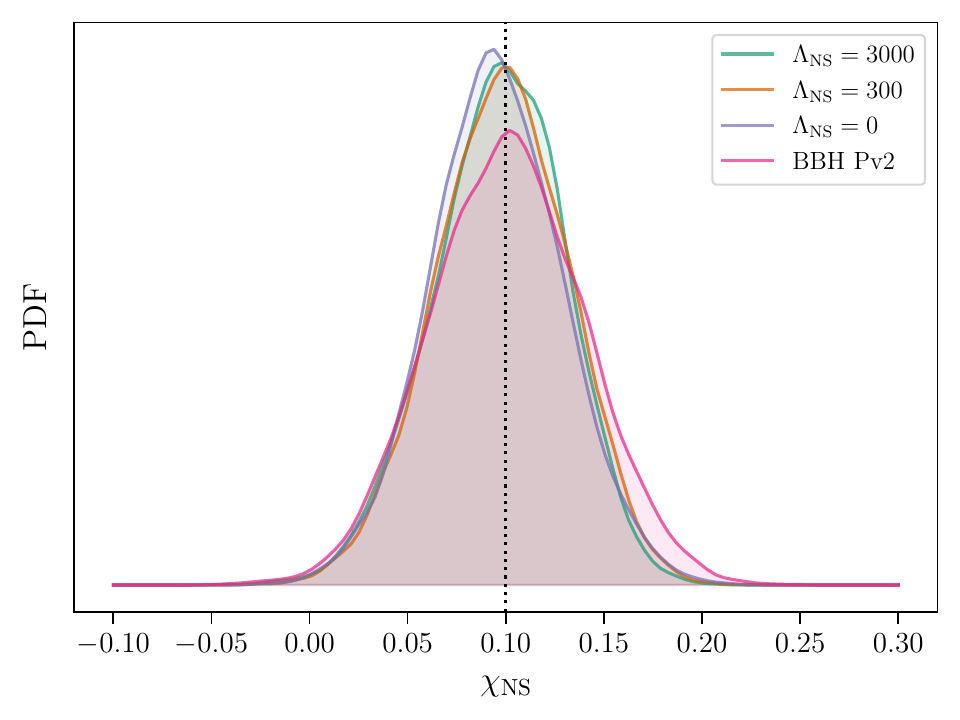}}
\caption{\label{fig:lambda} The inferred probability density function (PDF) for $\chiNS$ for NSBH systems with $(m_{\rm BH},m_{\rm NS}) = (5\msun,1.4\msun)$, at $\iota = 45^{\circ}$. The dotted vertical line shows the injected value of $\chiNS$.}
\end{figure}

During NSBH mergers, the NS can get tidally disrupted by the BH. The tidal disruption is affected by the equation of state of the NS, the spins of the components, and the mass ratio of the system (see Ref.~\cite{Kyutoku:2021icp} for a review). A high $\chiBH$ is conducive to the disruption, whereas a higher $m_{\rm BH}$ (hence, a low $q$) is unfavorable. Most of our estimates have been shown for NSBH systems with $m_{\rm BH} = 10\msun$ and $m_{\rm NS} = 1.4\msun$. For this system, even with high BH spins, we do not expect the NS to be significantly disrupted (e.g., Ref.~\cite{Kruger:2020gig}). However, for completeness, we show that the tidal contributions, especially through the spin-induced quadrupole moments, will not considerably affect $\chiNS$ estimates. To accomplish this, we choose IMRPhenomPv2\_NRTidalv2 \cite{Dietrich:2019kaq}, which contains the contribution of the tidal terms to the spin-induced quadrupole. Note that a more dedicated NSBH waveform like IMRPhenomNSBH \cite{Thompson:2020nei} was not used as it does not allow for spinning NS. We pair a $5\msun$ BH (which is the lightest BH considered in this study) with a $1.4\msun$ NS. The tidal deformability parameter for the BH is assumed to be 0, and for the NS, it is taken to be $\Lambda_{\rm NS} \in \{0, 300, 3000\}$. The $(\chiBH,\chiNS)$ where chosen to be $(0.8,0.1)$ to maximize the effect. We also compare these results with BBH keeping all parameters the same (except $\Lambda$, which would be 0 for a BH), and using the IMRPhenomPv2 \cite{Hannam:2013oca,Khan:2018fmp} waveform model. Note that while both IMRPhenomPv2 and IMRPhenomPv2\_NRTidalv2 allow for precession, we only consider non-precessing systems in this study. The corresponding plots are shown in figure~\ref{fig:lambda}. From the figure, we do not see a significant difference in the $\chiNS$ measurement as the value of $\Lambda_{\rm NS}$ is increased. The constraints obtained for the BBH waveform model are slightly broader, which may indicate that the inclusion of spin-induced quadrupole terms helps the $\chiNS$ measurement. However, a more detailed study needs to be done to make this assertion. For our purpose, we conclude including tidal effects will not hamper the $\chiNS$ measurement. 

\bibliography{bibliography}
\end{document}